\documentclass{vldb}
\usepackage{caption}
\usepackage{subcaption}
\usepackage{graphicx}
\usepackage{xcolor}
\usepackage{balance}  
\usepackage{url}
\usepackage{amsmath}
\usepackage{mathtools}
\usepackage{geometry}
\usepackage{array,relsize}
\usepackage{ragged2e}
\usepackage{algorithm}
\usepackage[noend]{algpseudocode}
\usepackage{float}
\usepackage{multirow}
\usepackage{todonotes}
\usepackage[export]{adjustbox}
\usepackage{soul}

\usepackage[T1]{fontenc}
\usepackage[utf8]{inputenc}
\usepackage[english]{babel}
\usepackage[numbers]{natbib}

\algdef{SE}[VARIABLES]{Variables}{EndVariables}
   {\algorithmicvariables}
   {\algorithmicend\ \algorithmicvariables}
\algnewcommand{\algorithmicvariables}{\textbf{global variables}}

\algnewcommand\algorithmicglobals{\textbf{Globals:}}
\algnewcommand\GLOBALS{\item[\algorithmicglobals]}
\algrenewcommand\algorithmicrequire{\textbf{Input: }}
\algrenewcommand\algorithmicensure{\textbf{Output: }}

\newcommand*{\Scale}[2][4]{\scalebox{#1}{$#2$}}%
\renewcommand{\footnotesize}{\scriptsize}

\newdef{definition}{Definition}
\newdef{exmp}{Example}
\newdef{lem}{Lemma}

\vldbTitle{Keyword Aware Influential Community Search in Large Attributed Graphs}
\vldbAuthors{Md. Saiful Islam, Mohammed Eunus Ali, Yong-Bin Kang, Timos Sellis, and Farhana Murtaza Choudhury}
\vldbDOI{https://doi.org/10.14778/xxxxxxx.xxxxxxx}
\vldbVolume{12}
\vldbNumber{xxx}
\vldbYear{2019}

\begin{document}


\title{Keyword Aware Influential Community Search in Large Attributed Graphs}



%
%
%
%

\numberofauthors{5} 

\author{
%
%
\alignauthor
Md. Saiful Islam\\
\affaddr{BUET, Bangladesh}\\
\email{saifulislam@cse.buet.ac.bd}
\alignauthor
Mohammed Eunus Ali\\
\affaddr{BUET, Bangladesh}\\
\email{eunus@cse.buet.ac.bd}
\alignauthor Yong-Bin Kang\\
\affaddr{Swinburne University of Technology, Australia}
\email{ykang@swin.edu.au}
\and 
\alignauthor Timos Sellis\\
\affaddr{Swinburne University of Technology, Australia}\\
\email{tsellis@swin.edu.au}
\alignauthor Farhana M. Choudhury\\
\affaddr{Melbourne University, Australia}
\email{fchoudhury@unimelb.edu.au}
}



\date{\today}

\maketitle

\begin{abstract}

 We introduce a novel keyword-aware influential community query ($KICQ$) that finds \emph{the most influential communities} from an attributed graph, where an influential community is defined as a closely connected group of vertices having some dominance over other groups of vertices with the expertise (a set of keywords) matching with the query terms (words or phrases). We first design the $KICQ$ that facilitates users to issue an influential CS query intuitively by using a set of query terms, and predicates (AND or OR). In this context, we propose a novel word-embedding based similarity model that enables \emph{semantic community search}, which substantially alleviates the limitations of \emph{exact keyword} based community search. Next, we propose a new influence measure for a community that considers both the cohesiveness and influence of the community and eliminates the need for specifying values of internal parameters of a network. Finally, we propose two efficient algorithms for searching influential communities in large attributed graphs. We present detailed experiments and a case study to demonstrate the effectiveness and efficiency of the proposed approaches.
\end{abstract}


\section{Introduction} \label{sec:introduction}
Communities serve as a basic structure for understanding the organization of many real-world networks or graphs. These networks include academic networks like DBLP, social networks like Facebook or Twitter, biological networks like protein-protein interactions, and many more. Finding communities from such large graphs has received significant attention in recent years due to its diverse practical applications that include event organization~\cite{sozio2010community}, friend recommendation~\cite{naruchitparames2011friend}, and e-commerce advertisement~\cite{kretz2008virtual}. Traditionally, community search (CS) on a large graph involves finding a community around a given query vertex that satisfies \emph{query parameters} like \emph{connectivity} and \emph{cohesiveness constraints}~\cite{fang2016effective, fang2017effective, huang2017attribute, sozio2010community}. For example, by using such techniques, one can find a community from the DBLP network for an author as a query, where the community should be a connected subgraph and each member should be connected to at least two other members in the community. More recent research works~\cite{li2015influential,li2017most}  have focused on \textit{finding influential communities} from a graph. The common goal of finding influential communities is to find a closely connected group of users (vertices) who have some dominance over other users in the graph in a particular domain. 

In this paper, we consider large attributed graphs where vertices (e.g., authors) are augmented with attributes (e.g., keywords) and propose a novel and efficient solution for finding influential communities that address the following gaps in the previous works:

\begin{figure}[t!]
\centering
\fontsize{6pt}{10pt}\selectfont
\def\svgwidth{4in}
\resizebox{0.48\textwidth}{!}{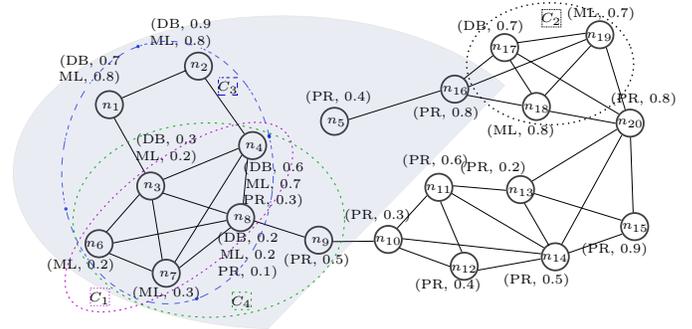}
\caption{ An attributed author-author graph, where each vertex has an associated list of attributes (keywords) and influences denoting her expertise. Different types cf communities are marked as $C_1-C_4$.}    
\label{fig:KSAN}
\vspace{-8mm}
\end{figure}

\textbf{First}, traditional CS works on an attributed graph require an input query vertex, and then find a group of neighboring vertices whose keywords have high similarity with the query vertex keywords. The resultant communities satisfy the required structural constraints~\cite{fang2016effective, huang2017attribute} (e.g., $C_1$ in Figure \ref{fig:KSAN} with $n_3$ as the query vertex, and parameter $k$ = 4 where $k$-truss is the structural constraint). A major limitation of such CS techniques is that the user needs to define the query vertex and the structural properties of the community explicitly, which might not be possible or suitable in many application domains.  A couple of recent studies~\cite{zhang2019keyword,chen2019contextual} tried to address these limitations by finding cohesive (i.e., $k$-core or triangle density) communities having close similarity with query keywords. However, they do not consider the influence of individuals in different keywords (e.g., $C_1$ in Figure \ref{fig:KSAN} is highly cohesive in terms of structure and keyword, but two highly influential vertices $n_1$, $n_2$ are ignored since influence is not considered) and also do not support flexible conjoining (using AND or OR predicates) of query keywords.



\textbf{Second}, existing works on influential community search only work on non-attributed graphs and also require specific values of structural parameters. For example, \cite{li2015influential} requires users to mention the value of $k$ while finding $k$-core based communities; similarly, \cite{li2017most} requires the values of the minimum number of vertices $m$ in a community and the maximum distance $p$ between any two vertices while finding an $mp$-clique based community. Although such parameters allow high customization in search, we argue that the choice of these parameters highly depends on the internal structure of the graph in practice. For example, if $k$ is set to a high value (e.g., 4) in a small graph (Figure \ref{fig:KSAN}), no community is returned by \cite{li2015influential} because there is no 4-core in this graph; and if $k$ is low (e.g., 2), the community returned (e.g., $C_2$ in Figure \ref{fig:KSAN}) does not have high cohesiveness. Similarly, given a query vertex, \cite{li2017most} only returns the desired community under specific constraints of parameter values (e.g., $C_4$ in Figure \ref{fig:KSAN} with parameters $m\le 6$ and $p = 2$), which is impractical for an external user. Thus flexibility in fixing parameters while searching for desired communities is crucial. 

\textbf{Third}, existing approaches to quantifying a community in terms of influence do not consider a comprehensive set of parameters that can affect the strength of a community. For example, the influence of a community is defined as the minimum influence among all members in \cite{li2015influential}; thus, a member with low influence can severely affect the influence of a community. We argue that an influence measure that considers both cohesiveness, the influence of individuals, and the size of the community should be considered while ranking communities, as all these factors contribute to the overall ranking of a community.

To fill the above research gaps and to support a new set of applications, we present a novel parameter free influential CS query, namely Top-$r$ Keyword-aware Influential Community Query ($KICQ$). To illustrate, let us consider an attributed graph of researchers, as shown in Figure \ref{fig:KSAN}. Here, vertices $n_1-n_{20}$ are authors who published papers in field of studies relevant to ``Machine Learning (ML)'', ``Database (DB)'', and ``Pattern Recognition (PR)''. An aspirant Ph.D. student may be interested in finding the most influential community who are working in ``ML'' or ``DB.'' The $KICQ$  returns the community $C_3$ as shown in Figure ~\ref{fig:KSAN} as the most influential community (see Section~\ref{sec:influence} for the influential metric)  since the members of the community have influence in either ``ML'' or ``DB'', the community is dense and also contains highly influential members.


The $KICQ$ can be useful for retrieving influential communities in other social networks, where a user is connected with her friends/followers, and her preferences are extracted from her posts/likes/shared contents. From such a network, one event organizer may want to find the most influential communities who have interests in ``music or movies''. Similarly, a tour operator may want to identify the groups who have ``travel'' as their interests as potential customers.

 A major challenge in realizing such a query (i.e., $KICQ$) comes from the fact that communities and their influences need to be computed and compared on the fly based on the set of keywords in the query, and thus existing pre-computation based approaches are not suitable for our purpose. Our major contributions are the following:

{\bf First}, we design $KICQ$ in such a way that enables users to issue an influential community search query intuitively by merely using a set of query terms (words or phrases), and predicates (AND or OR) (addressing the first limitation). In this context, we propose a novel word-embedding based keyword similarity model that enables \emph{semantic community search}, which substantially alleviates the limitations of \emph{exact keyword} based community search. For example, a user may use ``song'' instead of ``music,'' where exact search fails to retrieve the relevant communities if the attributed graph contains ``music'' as a keyword. \emph{This approach is of independent interest in enhancing any knowledge graph with semantically meaningful sets of words.} (Section~\ref{sec:augment_keywords})

{\bf Second}, we propose a new influence measure for a community that considers both the cohesiveness and influence of the community and eliminates the need for specifying values of internal parameters of a network (addressing the second limitation). The influence measure also captures the influence of individual members in a better intuitive sense rather than the influence of the community being dominated by the minimum influence of a member (addressing the third limitation). We demonstrate the effectiveness of the proposed measure in a case study. (Section~\ref{sec:influence})

{\bf Third}, we propose two efficient algorithms for searching influential communities in a large, attributed graph. The basis of the first algorithm is pruning the communities that cannot be a part of the answer set based on the computed scores of already explored subgraphs. The second one is a novel tree-based approach, where we augment the tree with influence score bounds for each keyword and prune the unnecessary branches of the tree based on the scores of the explored community. (Section~\ref{sec:algo})

{\bf Fourth}, we conduct comprehensive experiments with real datasets to evaluate our proposed algorithms. The experimental results show that our algorithms are highly efficient and effective in retrieving keyword aware influential communities compared to the state-of-the-art influential community search technique. (Section~\ref{sec:experiment})
\vspace{-2mm}

\section{Related Works}\label{sec:relatedworks}
Finding communities from large graphs has been an engaging research direction for a long time. Although the definition of community varies among different studies, cohesive subgraphs like maximal cliques~\cite{cheng2011finding}, $k$-core~\cite{cheng2011efficient}, $k$-truss~\cite{wang2012truss}, etc. form the basis of modeling communities. The task of finding communities can be divided into two major classes: community detection (CD), and community search (CS).  Recently, Li et al.~\cite{li2015influential} introduced the notion of \emph{influential community} that has piqued interest from the research community. CS and CD problems are studied on both simple and attributed graphs. 

 Team formation, another relevant domain, is the task of finding a subset of available individuals to complete a project which requires a specific set of skills, which can be viewed as a set coverage problem usually with a minimum communication cost objective~\cite{lappas2009finding, li2010team}. These works differ from CS as a  team does not need to be a cohesive subgraph.

\subsection{Community detection} \label{related/commDetection}
In CS, communities are defined based on the query, and CS solutions aim to find communities efficiently in an online manner. CD methods usually use global criteria to detect all the communities from an entire graph, where the focus is more on quality (e.g., cohesiveness) than efficiency. Link based analysis was popular in initial studies~\cite{fortunato2010community} that did not consider attributes in a graph. Clustering based techniques~\cite{zhou2009graph, ruan2013efficient, jiang2013efficient, xu2012model}, and topic modeling~\cite{liu2009topic, sachan2012using} are used in recent studies on attributed graphs. 
However, none of the studies enables a user to find specific communities of her interest, which is the main focus of our study.

\subsection{Community search} \label{related/commSearch}
We present different directions of CS studies in Table \ref{tab:related_cs}. Most of the basic CS studies on simple graphs ~\cite{cui2014local, huang2014querying, sozio2010community, barbieri2015efficient, yang2011social, yuan2017index} find communities containing given query vertices. Li et al.~\cite{li2018persistent} studied persistent communities in a temporal network, in which every edge is associated with a timestamp. Li et al.~\cite{li2015influential} introduced the notion of influential CS where vertices are assigned an influence score, and the influence of a community is modeled as the minimum influence of the members. Chen et al.~\cite{chen2016efficient} and Bi et al.~\cite{bi2018optimal} developed faster algorithms to solve the same problem. Zheng et al.~\cite{zheng2017querying} studied influential CS in an undirected weighted graph, where the weight of an edge represents the semantic intimacy between two vertices. Li et al.~\cite{li2017most} defined a community in terms of $kr$-clique and designed algorithms to retrieve the most influential community. All of these studies ignore rich information of vertices found in attributed graphs and require several vertices or internal parameters as part of a query, which is very difficult for a user who does not have enough knowledge of the graph.

\begin{table}[t!]
\centering
\small
\resizebox{\columnwidth}{!}{%
\begin{tabular}{|c|c|c|c|}
\hline
\multirow{2}{*}{CS Approaches} & \multirow{2}{*}{Simple graph} & \multicolumn{2}{c|}{Attributed graph} \\ \cline{3-4} 
                               &                               & Keyword   & Others   \\ \hline
Basic CS                       &    ~\cite{cui2014local, huang2014querying, sozio2010community, barbieri2015efficient, yang2011social, yuan2017index, li2018persistent}                           &  ~\cite{fang2016effective, huang2017attribute, chobe2019advancing, zhang2019keyword, khan2019compact, chen2019contextual}          &    ~\cite{chen2018exploring, fang2017effective, wang2018efficient, zhu2017geo}                       \\ \hline
Influential CS                 &    ~\cite{li2015influential, chen2016efficient, zheng2017querying, bi2018optimal, li2017most}                           &      -     &      ~\cite{li2018skyline}                     \\ \hline
\end{tabular}%
}
\caption{Existing community search works}
\label{tab:related_cs}
\vspace{-4mm}
\end{table}


There are several studies on CS in attributed graphs. Fang et al.~\cite{fang2016effective} proposed the ACQ algorithm to find subgraphs satisfying structural and keyword cohesiveness. Huang et al.~\cite{huang2017attribute} also explored attribute driven CS in terms of $k$-truss. Chen et al.~\cite{chen2018exploring} studied CS in an attributed graph where each vertex has a \emph{profile}: a set of keywords arranged in a tree structure. Chobe et al.~\cite{chobe2019advancing} employed keyword search techniques to facilitate CS in attributed graphs. However, these studies also require a set of vertices and/or internal parameters as part of the query. Few recent works~\cite{zhang2019keyword, chen2019contextual,zhang2019keyword} study keyword-based CS that take a set of keywords as input and return a subgraph as the community that has the \emph{best} match with the given set of query keywords. In these works, the cohesiveness of the subgraph is measured differently, i.e.,  k-core in~\cite{zhang2019keyword},  triangle density in~\cite{chen2019contextual}, and average proximity in~\cite{khan2019compact}. To decide a single best-matched subgraph, they define functions that consider the presence or absence of keywords and structural cohesiveness in the subgraph. These studies are different from ours as they only consider the presence or absence of keywords in different vertices of the subgraph and cannot be adapted for the scenario where we need to rank the communities and each vertex has a certain degree of influence in each keyword. There are CS studies on spatial graphs~\cite{fang2017effective, wang2018efficient, zhu2017geo} as well, which are of different interest to our problem.

Li et al.~\cite{li2018skyline} studies skyline community search where each vertex is associated with a $d$-dimensional influence score. However, their study is designed for low values of $d$ (i.e., $d<5$). With $d=5$, their algorithms require more than $10^3$ seconds in a graph with half a million vertices. This study cannot be extended for an attributed graph where vertices are associated with influence scores in multiple keywords, because there can be millions of keywords (dimension) in such an attributed graph. Also, this approach aims to find communities with a global objective function, and there is no way to search communities of specific interest.


\section{Problem Definition and System Overview}
\label{sec:overview}

We first define the attributed graph, proposed community model and keyword aware influential community query ($KICQ$), and then present the overview of the system.

\begin{definition}
(Attributed graph) An attributed graph \\$G^+(V, E, A)$ is an undirected graph, where $V$ is the set of vertices and $E$ is the set of edges. Each vertex $v$ is associated with a set of tuples of the form $A_v = \{(w_i, s_v(w_i))\}$, where $w_i$ is a keyword and $s_v(w_i) \in [0,1]$ is the influence score of vertex $v$ in keyword $w_i$.
\end{definition}

We consider the connected components of maximal $k$-cores as the \emph{influential communities}, where the \emph{influence} is defined as Equation~\ref{eq:comscore} (see Section~\ref{sec:influence}). $k$ is termed as \emph{cohesion factor} in this paper.

\begin{definition} \label{def:k-core}
(maximal $k$-core) Let $H$ be a subgraph of $G^+$, induced by the set of vertices $V_H \subseteq V$. Let the degree of a vertex $v$ in $H$ is denoted by $deg_H(v)$. $H$ is a $k$-core if $\forall_{v \in V_H} deg_H(v) \ge k$, where $k$ is a non-negative integer. $H$ is a maximal k-core if there is no super $k$-core in $G^+$ that contains $H$.
\end{definition}


Now, we define the keyword-aware influential community query, $KICQ$ as follows.

\begin{definition} \label{def:kicq}
($KICQ$) Let $G^+(V, E, A)$ be an attributed graph, $q(T,P)$ be a query tuple where $T = \{t_1, t_2, \cdots, t_n\}$ is a set of terms (i.e., words or phrases) and $P$ is a predicate (AND, OR) for conjoining the query terms, $k_{min}$ be the minimum cohesion factor of the resultant community, and $r$ be a positive integer specifying the number of top communities to be returned. Then, $KICQ$ finds $r$ \emph{most influential communities} $H_1,H_2,...,H_r$ from $G^+$. 
\end{definition}


\begin{figure}[!htb]
\centering
\includegraphics[trim = 180pt 170pt 420pt 112pt,width=0.48\linewidth]{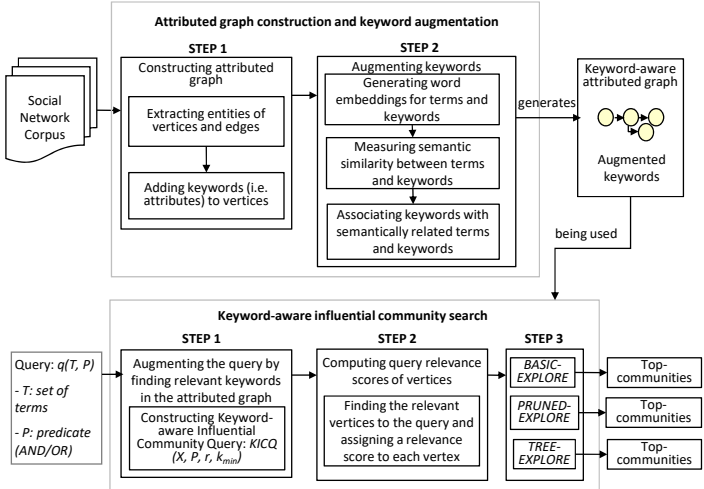}
\caption{The overview of our system}\label{fig:overview}
\vspace{-3mm}
\end{figure}

An overview of our system is presented in Figure \ref{fig:overview}. The system is mainly divided into two phases. First, we construct a keyword-aware attributed graph from a social network corpus that may consist of a combination of structured and/or unstructured (i.e., text) data. In an academic domain, the corpus can be scientific publications of researchers (e.g., authors, titles, abstracts, author-provided keywords, etc). Second, we focus on searching keyword-aware influential communities using the constructed attributed graph, given a query as a set of terms and predicates.

The distinctive features of the first phase are as follows: 
\begin{description}
    \item [STEP 1]: First, we build an attributed graph from a domain corpus by extracting entities of possible vertices and edges to represent the social network. 
    \item [STEP 2]: 
    To enable keyword-aware influential community search, we augment keywords with their semantically related terms and keywords.
    We build \textit{word embedding vectors}~\cite{Mikolov:2013} as an external knowledge source for associating keywords in the graph with semantically related terms and/or keywords. The semantic relatedness is estimated by exploiting the word embedding vectors. 
    The output of this step is a graph, called \textit{keyword-aware attributed graph}.

\end{description}
Further, the unique features of the second phase of our system can be briefly highlighted below:
\begin{description}
    \item [STEP 1]: Initially, a query raised by a user is given in the form of a pair $q(T, P)$ consisting of a set of query terms $T$, and a predicate $P$. The terms need to match with the keywords in the attributed graph to find meaningful communities. We acknowledge the difficulty faced by the users to put the exact terms while raising a query. For example, it is highly likely that some of the users will input “song” instead of  “music”. To help the users to easily raise a query, we augment each query term with a semantically meaningful set of keywords. The output of this step is a $KICQ$.
    \item [STEP 2]: Given a $KICQ$ query, our objective is to find the vertices relevant to the query and then compute their query relevance scores (Equation \ref{f:relevance}). Relevance score of vertices are used to compute the scores of potential influential communities. We argue that a measure that rewards both the cohesiveness of the community and high influence of the members, and does not require user input of any internal parameters (e.g., $k$ in $k$-core) is more preferable than the existing influence measures. We propose a linear weighted summation of the cohesiveness of the community and the total influence of the members of the community to estimate the overall score of a community (Section \ref{sec:influence}).
    \item [STEP 3]: Given the augmented query and the influential score function, our focus is now to retrieve top-$r$ most influential communities relevant to the query. Since we are the first to propose the keyword-aware influential community search problem, and existing pre-computation based approaches are not suitable to retrieve communities for any given query, we first present a basic solution named \texttt{BASIC-EXPLORE} followed by two efficient algorithms: \texttt{PRUNED-EXPLORE} and \texttt{TREE-EXPLORE}.
\end{description}


\section{Keyword augmentation}\label{sec:augment_keywords}

In a social network, one's expertise in various fields can be represented by a collection of terms (words or phrases). There are millions of such terms in a large network and it is difficult for users to come out with the exact terms while raising a query. We propose a semantic keyword similarity model that can augment any term with relevant keywords. This model is used to extend the keywords in the attributed graph, and associate appropriate keywords for each term in the query.

\subsection{Semantic keyword similarity model}\label{sec:sim_model}
Finding semantically related keywords of a term is not trivial. Basic preprocessing like removing whitespaces and stopwords are not sufficient. Two or multiple terms with a slight syntactic difference can indicate the same keyword (e.g., ``error detection and error correction'' and ``error detection and correction''). Even two different terms can represent the same keyword (e.g., ``AI'' and ``artificial intelligence'') or can be semantically similar (e.g., ``neural network'' and ``deep learning'').

We adopt Word2Vec~\cite{mikolov2013efficient} model to generate an embedding vector of any given word. We train this model with a domain corpus (e.g., scientific publications in an academic domain) after stopword removal, tokenization, and lemmatization~\cite{perkins2014python}.
A keyword/term can be thought of as a phrase containing one or many words. In our approach, the representative vector is formed using the average of the embedding vectors of the constituent words. 
Now, given any two terms $t_1$ and $t_2$, we denote their embedding vectors as \textbf{$x_{t_1}$} and \textbf{$x_{t_2}$}, respectively. To estimate a  similarity between these two terms, denoted as $S(t_1, t_2)$, we can use widely used \texttt{cosine} similarity of their embedding vectors $x_{t_1}$, $x_{t_2}$~\cite{muflikhah2009document}. We also propose a new similarity, which yields an attributed graph of better quality.\\ 
	\textbf{indirect cosine:} Given a term $t$, its vector $V^{t}$ is denoted as $V^{t} = [(w^{t}_1, s^{t}_1), (w^{t}_2, s^{t}_2), \cdots, (w^{t}_L, s^{t}_L)]$, where $w^t_i$ is the $i^{th}$ most similar word to $x_{t}$, $s^t_i$ is the corresponding similarity score, and $L$ is the number of similar terms of $t$.
	
	Given two terms $t_1$ and $t_2$, we construct a vocabulary, $U$ combining words of $V^{t_1}$, and $V^{t_2}$. Formally, $U = \{w: (w, s) \in V^{t_1} \cup (w, s) \in V^{t_2}\}$. To simplify our notation, let $U=\{w_1, w_2, \cdots, w_n\}$, where $n=|U|$. Now, we define another vector $SV^t = [s_1, s_2, \cdots, s_n]$, where $s_i$ is the similarity score of term $t$ to word $w_i \in U$, which can be found from $V^t$. If $(w_i, s_i) \notin V^t$, $s_i$ is set to 0. Finally, $S(t_1, t_2)$ is calculated as the cosine similarity of $SV^{t_1}$ and $SV^{t_2}$.

Finally, for any term $t$, we calculate its similarity with all the keywords in the given attributed graph, and find $M$-top most relevant keywords  $X_t$ ranked based on the similarity scores. $M$ is a system configurable parameter. By default, $M$ is set to 10.


\subsection{Constructing attributed graph}
In attributed graph $G^+$, a vertex $v$ is associated with a set of keywords. For all vertices, we extend each keyword $t$ with its $M$-top most relevant keywords $X_t$ using the semantic similarity model. For any keyword $w \in X_t$, the influence score of vertex $v$, $s_v(w) = s_v(t)$ since $w$ and $t$ are semantically similar.

\subsection{Augmenting keywords for KICQ}
A query $q(T, P)$ consists of a set of terms $T=\{t_1, t_2, \cdots,\\ t_n\}$ and a predicate $P$. First, the semantic similarity model is used to augment each term $t_i$ with the set of relevant keywords $X_{t_i}$. Then the system parameters $r$ and $k_{min}$ are used to formulate the keyword aware influential community query, $KICQ(X, P, r, k_{min})$ where $X=\{X_{t_1}, X_{t_2}, \cdots, X_{t_n}\}$.

\section{Influential Community Measures}
\label{sec:influence}

We design a scoring function that considers connectivity, cohesiveness, influence of individuals and the community size, and assigns a score for ranking the candidate communities given a query. 

First, for a given query, we redefine the influence of a vertex based on its relevance to the query. The query relevance score $\gamma_v$ of a vertex $v$ is estimated as follows: each vertex $v$ in the attributed graph is annotated with keywords and their influence score for the corresponding keywords, i.e., $(w_i, s_v(w_i))$. To estimate the relevance score, $\gamma_v \in [0,1]$, we need to consider the list of semantic keywords $X$ and the predicate $P$ in the $KICQ$ query. Formally, we use the following definition for computing $\gamma_v$:
\vspace{-2mm}
\begin{equation}\label{f:relevance}
    \gamma_v = \textbf{f}_{X_{t_i} \in X} [\textbf{g}_{w \in X_{t_i}} s_v(w)]    
\end{equation}

Here, $\textbf{f}$ and $\textbf{g}$ are two aggregate functions: $\textbf{g}$ combines the relevance of the vertex for the semantic keywords of a query term, and  $\textbf{f}$ combines the relevance scores in all terms considering the predicate $P$. We use $\textbf{g}$ as the aggregate function, \emph{MAX}; whereas we use $\textbf{f}$ as \emph{MIN} for AND predicate and \emph{MAX} for OR predicate, respectively. \emph{MIN} aggregate ensures that a vertex has high relevance to all the terms, while \emph{MAX} only requires high relevance to any of the terms. 

Now, we use a linear weighted summation of the cohesiveness and influences to calculate the overall score of a community. Let $H = (V_H,E_H)$ is a subgraph of attributed graph $G^+(V,E,A)$. If $H$ is a community (connected component of maximal $k$-core), then the score of $H$ is:
    \begin{equation}\label{eq:comscore}
    \zeta(H) = \beta \times \underbrace{\textstyle \frac{ k}{ max\text{-}deg(G^+)}}_{\text{Cohesiveness score}}  +  (1-\beta) \times \underbrace{\textstyle \frac{ \sum_{v \in V_H} { \gamma_v}  }{ |V|}}_{\text{Influence score}}
    \end{equation}
Here, $max\text{-}deg(G^+)$ is the maximum degree of all vertices in $G^+$. Both the cohesiveness and the influence score of a community are normalized within $[0,1]$, and the preference parameter $\beta \in [0,1]$ defines the importance of one score relative to the other. 

Since we model a community using connected $k$-core, connectivity and cohesiveness is ensured. $max\text{-}deg(G^+)$ and $|V|$ is constant for attributed graph $G^+$. Thus the influence score of community $H$ depends on $\sum_{v \in V_H} { \gamma_v}$ which prefers large community with highly influential individuals.  Also as long as some low influential members do not disrupt the cohesiveness of the community, the score of this community is not penalized, which is the case in~\cite{li2015influential}. We acknowledge that such a measure is not unique, and other measures can be explored in the future. However, experiments using real datasets and the case study presented in Section \ref{sec:caseStudy} demonstrate that our proposed influence measure can capture cohesive communities with highly influential members.

\section{Algorithms for Influential Community Search } \label{sec:algo}


In this section, we present algorithms for finding $r$ most influential communities from the attributed graph $G^+$ for a given $KICQ(X, P, r, k_{min})$. Since the notion of \emph{influential community} changes with different sets of query keywords, existing pre-computation based approach~\cite{li2015influential, li2017most} cannot be adapted for this purpose. 

\subsection{A straightforward approach, \texttt{BASIC-EXPLORE}} \label{sec:icr_basic}

A straightforward approach to answer $KICQ$ on a large graph is as follows. First, we extract the subgraph, which we call the \emph{query essential subgraph}, $G_q$, containing vertices and edges that are relevant to the query. Then we find all the connected components of maximal $k$-core subgraphs for all possible values of $k$. Finally, we return the top $r$ communities having the highest influential community scores as per Equation~\ref{eq:comscore}.


\textbf{\emph{Finding $G_q$.}} The query essential subgraph, $G_q(V_q, E_q, \gamma)$ is a subgraph of the attributed graph $G^+(V, E, A)$ induced by $V_q$, the set of vertices with a non-zero query relevance score. In $G_q$, each vertex $v$ is annotated with its relevance score $\gamma_v$, and $E_q$ is the set of edges between any two vertices in $V_q$. To efficiently generate the $G_q$, we maintain an inverted index, where for each keyword $w$, a list $IL_w$ of the vertices that contain $w$ is stored. Thus, for a given $KICQ$ query, $V_q$ can be obtained by,

\begin{equation}
    V_q = 
    \begin{cases}
    		\bigcap_{X_{t_i} \in X} [\bigcup_{w \in X_{t_i}} IL_w],& \text{if } P = AND\\
    		\bigcup_{X_{t_i} \in X} [\bigcup_{w \in X_{t_i}} IL_w],              & \text{otherwise}
	\end{cases}
	\label{eq:qeg-vertices}
\end{equation}

After retrieving $V_q$, we compute the query relevance score of each vertex $v \in V_q$ (Equation \ref{f:relevance}) and retrieve $E_q$ that denotes the connections between all pairs of vertices in $V_q$.




\textbf{\emph{Finding $k$-cores and most influential communities.}} Algorithm \ref{alg:basicExplore} outlines the procedure \Call{basic-explore}{} for finding $r$ most influential communities from $G_q$. First, we compute core decomposition for all vertices in $G_q$ using the $O(|E_q|)$ algorithm proposed by Batagelj et al.~\cite{batagelj2003m}. A priority queue $Q$ is used to hold our solution. We initialize $Q$ with $r$ empty communities having 0 score. In our case, a community must be at least $k_{min}$-core. Again, the maximum cohesion factor of a community in $G_q$ can be $max\text{-}deg(G_q)$, since there is no vertex in $G_q$ with a higher degree. Thus we need to first find all connected components of maximal $k$-cores from $G_q$, where  the value of $k$ is in range $[k_{min}, max\text{-}deg(G_q)]$. Then, we compute the influential scores of each computed community, and finally, maintain the top-$r$ communities in $Q$ ordered by the scores of the communities.

\setlength{\textfloatsep}{4pt}
\begin{algorithm}[!t]
    \begin{smaller}
        \caption{BASIC-EXPLORE ($G_q$)}
        \label{alg:basicExplore}
        \begin{algorithmic}[1]
            \State compute core decomposition for all vertices in $G_q$
            \State initialize a priority queue $Q$ with $r$ empty communities (score 0)
            
            \For{$k=k_{min}$ to $max\text{-}deg(G_q)$}
                \State $H^{k}=$ maximal $k$-core in $G_q$
                \State $CC^{k}=$ set of connected components in $H^{k}$
                
                \For{all $h(V_h, E_h) \in CC^{k}$}
                    \State $\zeta(h) = \text{score of } h$
                    
                    \If{$\zeta(h)>r^{th}\text{ best score}$}
                    \State $Q$.pop()
                    \State $Q$.push($h$) 
                    \EndIf
                    
                \EndFor
                
            \EndFor
            
        \end{algorithmic}
	\end{smaller}
\end{algorithm}

\textbf{\emph{Time complexity.}} Finding the relevant vertices and calculating their relevance score can be done in $O(|V_q| \times N_w)$ time, where $N_w = \sum_{X_i \in X}|X_i|$ is the total number of relevant keywords. If the graph is implemented with adjacency list, $E_q$ can be obtained in $O(|V_q|)$ time by taking union of adjacency list and $V_q$ for each vertex.
So, time complexity for computing $G_q$ is $O(|V_q| \times N_w)$. Core decomposition of $V_q$ is done in $O(|E_q|)$ time. The operations (push, pop) performed in the priority queue takes $O(\log r)$ time. So, the time required for initializing $Q$ is $O(r\log r)$. Exploring a maximal $k$-core requires computing its connected components ($O(|V_q|+|E_q|)$), obtaining $k$-core vertices ($O(|V_q|)$), computing scores of each connected components, and updating the priority queue $Q$. For any community $h(V_h, E_h)$, the run-time for computing its score is bounded by $O(|V_h|) =  O(|V_q|)$ (simplified). So, if $N_k$ is the number of influential communities with cohesion factor $k$, then the runtime of exploring all $k$-cores is bounded by $O(max\text{-}deg(G_q) \times ((|V_q|+|E_q|) + N_k \times (|V_q|+log(r)))$. Considering $|V_q|>log(r)$, the bound can be simplified as $O(max\text{-}deg(G_q) \times |V_q|^2)$ for a dense graph \footnote{$N_k < |V_q|$ and for any dense graph $G(V,E)$, $|E|$ is $O(|V|^2)$}. Since, this dominates the time complexity of finding $G_q$, we can conclude that, the overall complexity of \texttt{BASIC-EXPLORE} is \textbf{$O(max\text{-}deg(G_q) \times |V_q|^2)$}.
\subsection{Pruned exploration approach, \texttt{PRUNED-EXPLORE}} \label{sec:icr_prune}

The major bottleneck of \texttt{BASIC-EXPLORE} is that it needs to explore all maximal $k$-cores, for different values of $k$, and find the connected components of each maximal $k$-core subgraph. Such exploration is computationally expensive for a large graph. 
Instead of directly exploring the subgraphs to compute the maximal $k$-core and its connected components (communities), we first estimate the upper bound score of the communities of the corresponding subgraph. This bound can be used to prune a large number of redundant subgraphs that cannot be a part of the top-$r$ influential communities.

	First, we find the query essential subgraph $G_q$, compute core decomposition, and initialize priority queue $Q$ as described in Section \ref{sec:icr_basic}. Now, we need to explore $G_q$ to retrieve communities for all possible values of $k$. As discussed before, the value of $k$ must be between $k_{min}$  and $max\text{-}deg(G_q)$. We propose the following lemmas, which pave the foundation of our pruning.

    \begin{lem} \label{lemma:noExpand}
        Let, $H(V_H, E_H)$ be a subgraph of $G_q$. 
        For any community in $H$, the maximum influence score can be the sum of the query relevance scores of all vertices in $H$. Thus, without computing the vertices of $k$-core subgraph, we can calculate the upper bound of the score of any community in $H$ for a particular value of $k$ as follows.
        
        \begin{equation}\label{eq:upperscore}
            \zeta^*_k(H) = \beta \times \textstyle \frac{ k}{ max\text{-}deg(G^+)}  +  (1-\beta) \times \textstyle \frac{ \sum_{v \in V_H} { \gamma_v}  }{ |V|}
        \end{equation}
    \end{lem}
    
    \begin{lem} \label{lemma:D_min}
        If $H = (V_H,E_H)$ is a subgraph and $min\text{-}deg(H) =  \min\limits_{v \in  V_H} (deg_{H}(v)) > k$, then any community in $H$ must be at least $min\text{-}deg(H)$-core.
    \end{lem}
    
According to Lemma~\ref{lemma:noExpand}, we can prune a subgraph if its upper bound score is lower than the $r^{th}$ best score of already retrieved communities from $G_q$. Moreover, Lemma~\ref{lemma:D_min} helps us to avoid the computation of certain cores from $G_q$.

  Now, we develop a recursive procedure \Call{pruned-explore}{} to search for influential communities in $G_q$. Algorithm \ref{pruneAndExplore} outlines the procedure. Initially, \Call{pruned-explore}{$G_q,k_{min}$} is called to extract communities with minimum cohesion factor. In later steps, the procedure is recursively called to extract communities with higher cohesion factors. 
  
    \begin{algorithm}[!t]
        \begin{smaller}
        	\caption{PRUNED-EXPLORE ($H$, $k$)}\label{pruneAndExplore}
            \begin{algorithmic}[1]
                \Require Subgraph of $G_q$ $H=(V_{H},E_{H})$, cohesion factor $k$.
        	    
                \State $min\text{-}deg(H)=\min\limits_{v \epsilon V_{H}} (deg_{H}(v))$
                \If{$min\text{-}deg(H)>k$}
                    \State $k=min\text{-}deg(H)$
                \EndIf
                \State $H^{k}=$ maximal $k$-core in $H$
                \State $CC^{k}=$ set of connected components of $H^{k}$
                \For{all $h \in CC^{k}$}
                    \If{actual score, $\zeta(h)>r^{th}\text{ best score}$}
                        \State $Q$.pop()
                        \State $Q$.push($h$)
                    \EndIf
                    \For{$k^\prime=k+1$ to $max\text{-}deg(G_q)$}

                        \If{upper bound score, $\zeta^*_{k^\prime}(h)>r^{th}\text{ best score}$}
                            \State{\Call{PRUNED-EXPLORE}{$h,k^\prime$}}
                            \State{break}
                        \EndIf
                    \EndFor
                        
                \EndFor
            \end{algorithmic}
        	
	    \end{smaller}
        \end{algorithm}

Let us consider that we want to find communities with cohesion factor $k$, from a subgraph $H(V_H, E_H)$ of $G_q$. In lines 1-3, we determine the minimum degree of the vertices in $H$, $min\text{-}deg(H)$. If $min\text{-}deg(H) > k$, we set $k = min\text{-}deg(H)$ and directly compute such $k$-cores (according to Lemma \ref{lemma:D_min}). In lines 4-5, we find the set of connected components of maximal $k$ core of $H$, denoted by $CC^k$. The loop in line 6 runs for each connected component. We update the priority queue if any connected component's score is higher than the current top-$r$ communities in lines 7-9. We explore the connected component for higher values of $k$ in lines 10-13. We use Lemma \ref{lemma:noExpand} to prune exploration for the values of $k$ for which the upper bound of the score is lower than the $r^{th}$ best community. When the procedure terminates, the queue holds the final top-$r$ communities.
\vspace{-2mm}
\subsection{Keyword indexed tree exploration, \texttt{TREE-EXPLORE}} \label{sec:icr_scalable}

Though the above \texttt{PRUNED-EXPLORE} can prune a large number of subgraphs based on the derived upper bounds, it still explores subgraphs and their connected components with low cohesiveness, which usually do not contain the most influential communities. This exploration can be costly, especially in a scenario where the query essential graph, $G_q$, turns out to be very large. So, we propose a novel index, namely \emph{keyword indexed core-label tree} (\texttt{KIC-tree}), that pre-computes and organizes the connected components of maximal $k$-core subgraphs hierarchically with computed upper bound of influence scores for each keyword.

The key idea of our \texttt{KIC-tree} based $KICQ$ comes from the following observations: 

(i) Top communities are structurally cohesive and thereby can be retrieved by exploring the subgraphs of higher cohesion factors. Thus, if $k$-cores are precomputed, disregarding the associated keywords, we can still prune the subgraphs with low $k$ value.

(ii) Communities are represented using connected maximal $k$-cores which are nested, i.e., by definition, a $(k+1)$-core is also a $k$-core ($k \ge 0$). This property helps to store all the connected components of maximal $k$-cores in compressed tree-based structures as shown in previous works \texttt{ICP-index}~\cite{li2015influential}, \texttt{CL-tree index}~\cite{fang2016effective}.

(iii) We can compute the upper bounds for both the components: influence and cohesiveness, of the the scoring function, and use these upper bounds to prune the search space during query time.


We first discuss the basic structure of the \texttt{KIC-tree} index. Then we present the upper bounds for individual keywords and aggregate them for a set of keywords and predicates (in $KICQ$) for an upper bound score of a node. We also show how the cohesiveness score can be bounded based on a pre-computed structure alone. Finally, we present our \texttt{TREE-EXPLORE} algorithm for influential community search using the \texttt{KIC-tree}. In this section, we use the term ``node'' to exclusively indicate a tree node.

\subsubsection{KIC-tree index} \label{sec:basic-cl-tree}
The \texttt{KIC-tree} index organizes the connected components of $k$-cores into a space-efficient tree structure. We adopt the concept of compressed tree based structure  of previous works (e.g., \texttt{CL-tree index}~\cite{fang2016effective}), and augment the structure with derived bounds to prune the search space. 



\begin{figure}[t!]
\centering
\fontsize{8pt}{12pt}\selectfont
\def\svgwidth{4in}
\resizebox{0.48\textwidth}{!}{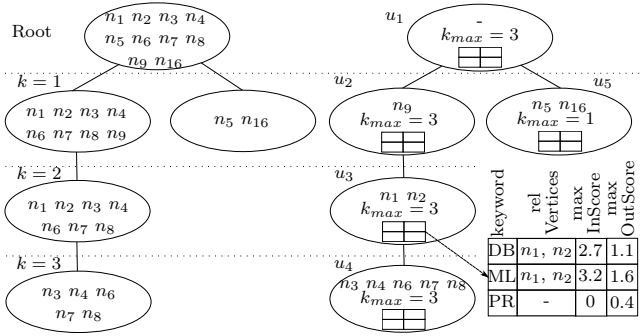}
\caption{KIC-tree for the subgraph (shaded) in Figure \ref{fig:KSAN}.}    
\label{fig:cltree}
\vspace{-1mm}
\end{figure}

Figure \ref{fig:cltree} shows an example \texttt{KIC-tree} for the subgraph shown as the shaded region in Figure \ref{fig:KSAN}. The left shows the hierarchical representation of all maximal k-core  connected components in the subgraph. We refer this tree as the \emph{uncompressed tree}. The right figure shows \texttt{KIC-tree} index, a more compact representation of the left tree, which removes the graph vertices present in its descendant nodes ensuring that each graph vertex appears exactly once. 




Let $u$ be a \texttt{KIC-tree} node and $subtree(u)$ be the subtree rooted at $u$. The structure of $u$ is as follows:

(i) $k$, the cohesion factor;
(ii) $vertexSet$, the set of compressed graph vertices at node $u$;
(iii) $childNodes$, the set of child nodes of $u$;
(iv) $k_{max}$, the maximum cohesion factor of any connected component contained by the subtree($u$);
(v)  $iList$, an inverted list containing the upper bounds of influence scores for all keywords appeared in subtree($u$).


For each keyword $w$ that appears in subtree($u$), the inverted list $u.iList[w]$ contain the following elements:

(i) $relV$, a set of graph vertices in $u.vertexSet$ containing the keyword $w$;

(ii) $maxKNScore$, the upper bound of influence score component  by only considering keyword $w$ in a community (i.e., a connected component) contained by the $subtree(u)$, where the community must include at least one vertex present in node $u$ containing the keyword $w$;

(iii) $maxKDScore$, the upper bound of influence score component  by only considering keyword $w$ of a community contained by the subtree($u$), where the community does not include any vertex from $u.vertexSet$ (i.e., all vertices of the community come from the descendent nodes of $u$).

We compute $maxKNScore$ and $maxKDScore$ as follows. 

For a node $u$, let $childV$ be the set of graph vertices stored at the descendent nodes of $u$, and $allV$ be the set of graph vertices at $subtree(u)$ (i.e., all the vertices in node $u$ and its descendent nodes). If $u$ is a leaf node, $u.childV = \emptyset$. Otherwise, $u.childV = \bigcup_{p \in u.childNodes}p.vertexSet$. On the other hand, in all cases, $u.allV = u.vertexSet \bigcup u.childV$.

Now, if there is no relevant graph vertex in node $u$ for keyword $w$, then we set $maxKNScore$ as 0. Otherwise, the upper bound is the sum of influence scores of all graph vertices in $u.allV$. Formally,
    $$
    \Scale[0.90]{
        u.iList[w].maxKNScore =  
            \begin{cases}
        		0, \text{if } u.iList[w].relV = \emptyset\\
        		\sum_{v \in u.allV}(s_v(w)),              \text{otherwise}
    	    \end{cases}
    }
    $$
Here, $s_v(w)$ is the influence score of vertex $v$ for keyword $w$.

Now, $maxKDScore$ is the maximum influence score component among the communities represented by the descendant nodes of $u$. $u.iList[w].maxKDScore = 0$ if $u$ is a leaf node. Otherwise, we can use the computed values of $maxKNScore$ to compute the $maxKDScore$ as follows.
    $$
        \Scale[0.85]{u.iList[w].maxKDScore =  \max_{p \in u.childNodes} p.iList[w].maxKNScore}
    $$
Figure \ref{fig:cltree} (right) shows an example tree, where the table inside the ellipse represents the $iList$ of the corresponding node. For simplicity, we only show the $iList$ for node $u_3$.

\subsubsection{Complexity analysis for index construction:} \label{sec:cltree_space}
We use the \texttt{advanced} method proposed by Fang et. al. \cite{fang2016effective} that compresses the tree and for each node $u$, computes $u.iList[w].relV$ for all the relevant keywords of $u$. The time complexity of this method is $O(|E| \times \alpha(|V|))$, where $\alpha(|V|)$, the \texttt{inverse Ackermann function}, is less than 5 for all remotely practical values of $|V|$. For each $iList[w]$ entry, we also need to compute the two upper bounds $maxKNScore$ and $maxKDScore$. If $A_{max}$ is the maximum number of keywords associated with a graph vertex, the time complexity for computing $maxKNScore$ is $O(A_{max}\times|V|)$. Computing $maxKDScore$ for a node $u$ only requires visiting its $childNodes$ which is non-dominant.
So, overall time complexity for index construction is $O(|E| \times \alpha(|V|) + A_{max}\times|V|)$.

In $iList$, we need additional space to store two upper bound scores (constant space) for each keyword. The space cost is still dominated by storing $relVertices$ in $iList$. So, the space complexity remains $O(\bar{A} \times |V|)$ as in ~\cite{fang2016effective}, which is proportional to the graph size.

\subsubsection{Computing upper bound scores for a query}
Given a $KICQ(X, P, r, k_{min})$ query, we need to compute an upper bound influence score of a community denoted by $S_{inf}$ and the maximum possible cohesiveness score of that community, $S_{k}$ by using the precomputed upper bounds in \texttt{KIC-tree}. Then the upper bound of the total score of that community can be computed as $maxScore = \beta \times S_k + (1-\beta) \times S_{inf}$ (as in Equation \ref{eq:comscore}).


We define two upper bounds for the communities inside $subtree(u)$: (i) $maxNodeScore$, the maximum possible score of any community that can be exclusively found by exploring the connected $k$-core stored at node $u$ and (ii) $maxDesScore$, the maximum possible score of any community that can be found by exploring the descendant nodes of $u$.

\textbf{Computing $maxNodeScore$:}
 For any community contained exclusively by node $u$, there must be at least one vertex $v$ that is stored  at $u$. Now, for any vertex $v$ exclusive to node $u$, $u.k$ is the maximum core number. So, a subgraph containing $v$ can be at most $u.k$-core (irrespective of any keyword) and the upper bound of cohesiveness score of any community contained by the node can be computed as $S_{k} = u.k/max\text{-}deg(G^+)$. 

Now, for each keyword $w$, $u.iList[w].maxKNScore$ already defines the upper bound of influence score component for any community in the subgraph exclusively contained by node $u$ (considering the single keyword $w$). We combine these bounds for considering all the keywords in the $KICQ$ query and compute the maximum influence score as: 

$S_{inf} = \frac{1}{|V|} \times \textbf{F}_{X_{t_i} \in X}(\sum_{w \in X_{t_i}} u.iList[w].maxKNScore)$ 

Here $\textbf{F}$ is an aggregate function that combines the influence scores of the community for multiple terms depending on the predicate $P$ and division by $|V|$ normalizes the score within [0,1]. For the queries with OR predicate, a top community can be formed by joining multiple communities pre-computed for a single term, and these communities may have disjoint vertex set. So, it is safe to consider $\textbf{F}$ as a \textbf{summation} aggregate. For the same reason, $\sum$ is explicitly used to combine the semantic keywords of a term. Again, for the queries with AND predicate, any graph vertex forming a community for a single term must be present in communities of other terms as well. So, $\textbf{F}$ can be safely considered as \textbf{minimum} aggregate.

\textbf{Computing $maxDesScore$:}
For any community contained by the descendant nodes of $u$, the maximum cohesion factor is $u.k_{max}$ and the upper bound of cohesiveness score is $S_{k} = u.k_{max}/max\text{-}deg(G^+)$. 

Again, for keyword $w$, $u.iList[w].maxKDScore$ already defines the upper bound of influence score component for any community contained by the descendant nodes. We combine these bounds for considering all the keywords in the $KICQ$ query and compute the maximum influence score similarly as computing $maxNodeScore$, i.e.,

$S_{inf} = \frac{1}{|V|} \times \textbf{F}_{X_{t_i} \in X}(\sum_{w \in X_{t_i}} u.iList[w].maxKDScore)$

\subsubsection{\texttt{TREE-EXPLORE} algorithm}
We follow a \emph{best-first} exploration strategy. Since the leaf nodes contain the communities with high cohesiveness while nodes near root contain communities with low cohesiveness, we explore the \texttt{KIC-tree} in a post-order manner. Likewise the previous exploration algorithms (e.g., \texttt{PRUNED-EXPLORE}), a priority queue $Q$ initialized with $r$ empty communities is used to store the results. The exploration algorithm, which we call \texttt{TREE-EXPLORE} is developed based on the following pruning techniques:

\textbf{(i) Subtree pruning}: For any node $u$, we examine the $u.maxDesScore$ before visiting its children. If it is less than the $r^{th}$ best score, then none of the communities to be found in the descendent nodes can score higher than the current $r^{th}$ top community. Therefore, we can skip visiting the descendant nodes of $u$.

\textbf{(ii) Node pruning}: Before exploring the pre-computed connected $k$-core subgraph at any node $u$, we examine the $u.maxNodeScore$. If it is less than the $r^{th}$ best score, we can safely prune this exploration.

Algorithm \ref{alg:cltreevisit} outlines the pseudocode for the \texttt{KIC-tree} traversal. The inverted list that we have used to find the $G_q$ is adopted for computing $U$, the set of tree nodes relevant to a query. Initially, the recursive procedure \Call{tree-explore}{u, $U$} is called with $u$ being the root of the \texttt{KIC-tree}. 

\begin{algorithm}[!t]
        \begin{smaller}
        	\caption{TREE-EXPLORE ($u, U$)}\label{alg:cltreevisit}
            \begin{algorithmic}[1]
                \Require Tree node $u$, query relevant nodes $U$.
                \If{$u$ is an internal node}
                    \State Compute influence score component $S_{inf}$ and cohesiveness score component $S_{k}$ for $u.maxDesScore$
                    \State $u.maxDesScore = \beta \times S_{k} + (1-\beta) \times S_{inf}$
                    \If{$S_{inf}>0$ and $u.maxDesScore> r^{th}\text{ best score}$}
                        \For{each $p \in (u.childNodes \cap U)$}
                            \State{\Call{TREE-EXPLORE}{$p, U$}}
                        \EndFor
                    \EndIf
                \EndIf
                
                \If{$u.k < k_{min}$}
                    \State return
                \EndIf
                
                
                \State Compute influence score component $S_{inf}$ and cohesiveness score component $S_{k}$ for $u.maxNodeScore$
                \State $u.maxNodeScore = \beta \times S_{k} + (1-\beta) \times S_{inf}$
                \If{$S_{inf}=0$ or $u.maxNodeScore< r^{th}\text{ best score}$}
                    \State return
                \EndIf
                                
                \State $u.V_{rel} = $ compute relevant graph vertices in the subtree rooted at $u$
                \State Compute query relevance score of all vertices in $u.V_{rel}$
                \State Compute $u.E_{rel}$, the edges among $u.V_{rel}$
                \State Construct subgraph $H(u.V_{rel}, u.E_{rel})$, each vertex annotated with relevance score
                \State{\Call{modified-pruned-explore}{$H$, $k_{min}$, $u.k$}}
                    
            \end{algorithmic}
        	
	    \end{smaller}
        \end{algorithm}
        
For any internal node $u$, we first compute and examine the influence score component, $S_{inf}$, and the cohesiveness score component, $S_k$ of
$u.maxDesScore$. If $S_{inf}$ is 0, the descendants of $u$ do not contain any graph vertex relevant to the query, and therefore we do not need to visit subsequent nodes across the subtree. If $S_{inf}>0$ and $u.maxDesScore$ is greater than the current $r^{th}$ best score, then we visit its children (lines 1-6).

Now we want to explore the pre-computed connected $k-core$ subgraph represented by node $u$. If the cohesion factor $k$ in node $u$ is less than $k_{min}$, then we prune exploring the subgraph. Otherwise, we compute the influence score component ($S_{inf}$) and the cohesiveness score component ($S_{k}$) of $u.maxNodeScore$. If $S_{inf}$ is 0, then the node does not contain any graph vertex relevant to the query, and we can safely skip exploring the subgraph. Again, we skip the exploration if $u.maxNodeScore$ is less than the current $r^{th}$ best score (lines 7-12). 

If the exploration of the connected $k$-core subgraph cannot be pruned, we first need to find all the relevant graph vertices, $u.V_{rel}$ present in the subgraph (line 13). Since \texttt{KIC-tree} compresses these graph vertices by removing ones present at descendant nodes, we need to decompress in a bottom-up manner. At any node $u$, the relevant graph vertices $u.V_{rel}$ can be computed like the vertices in $QEG$ (Equation \ref{eq:qeg-vertices}) just by replacing $IL_w$ with $u.iList[w].relV$.  For any internal node $u$, we need to add the relevant graph vertices in child nodes to $u.V_{rel}$.

Now we compute the relevance score of each vertex $v \in u.V_{rel}$ (Equation \ref{f:relevance}) and then compute the edges among these vertices, thereby construct the subgraph $H(u.V_{rel}, u.E_{rel})$ (lines 14-16).

The procedure {\small \Call{modified-pruned-explore}{$H, k, k_{max}$}} is a slightly modified version of the procedure {\small \Call{pruned-explore}{$H, k$}} that takes an extra argument $k_{max}$, the maximum value of cohesion factor for sub-graph $H$. Lines 10-13 in Algorithm \ref{pruneAndExplore} are replaced by the following:\\
\vspace{-2mm}
    \begin{small}
    \begin{algorithmic}[1]
    \setcounter{ALG@line}{9}
    
        \For{$k^\prime=k+1$ to $k_{max}$}
            \If{$\zeta^*_{k^\prime}(h)>r^{th}\text{ best score}$}
                \State{\Call{modified-pruned-explore}{$h,k^\prime, k_{max}$}}
                \State{break}
            \EndIf
        \EndFor
                        
    \end{algorithmic}
    \end{small}
    
Since no graph vertex at $u$ belongs to any $k$-core with cohesion factor higher than $u.k$, here $k_{max} = u.k$. Initially, $k = k_{min}$. So, \Call{modified-pruned-explore}{$H, k_{min}, u.k$} is called to explore the subgraph $H$ (line 17).


\section{Experimental Study} \label{sec:experiment}
In this section we present experiments to evaluate the performance of our proposed algorithms. 

\subsection{Experimental setup} \label{sec:setup}
All the community search algorithms are implemented in JAVA. Experiments were run on a virtual environment of OzSTAR\footnote{https://supercomputing.swin.edu.au/ozstar/} supercomputer with two cores of Intel Gold 6140 CPU @ 2.30 GHz 2.30GHz, 192 GB RAM, and 400 GB SSD. We assume that the graph and all the indexes will fit in the memory. 
For the simplicity of presentation, we use shorter names for our algorithms: \texttt{BASIC}, \texttt{PRUNE}, and \texttt{TREE} to represent \texttt{BASIC-EXPLORE}, \texttt{PRUNED-EXPLORE}, and \texttt{TREE-EXPLORE} respectively. We present the average results of 100 queries.

We use two large real datasets: \texttt{OAG} (Open Academic Graph)\footnote{https://aminer.org/open-academic-graph} \cite{tang2008arnetminer} and \texttt{Gowalla}\footnote{http://www.yongliu.org/datasets/index.html} that reflect the real life application scenarios. In \texttt{OAG} dataset, we represent first 1 million authors as vertices and $15,677,940$ co-authorship relations among the authors as edges. We choose $1,000$ most frequent author-provided keywords as the set of keywords for the attributed graph and then apply our \texttt{semantic similarity model} to extend the keywords as mentioned in Section \ref{sec:sim_model}. In \texttt{Gowalla} dataset, users and friendship among them are modeled as vertices and edges, and the location ids are considered as keywords. There are $407,533$ vertices, $2,209,169$ edges, and $2,727,464$ keywords in this attributed graph. For both datasets, the influence score of a user for a certain keyword is modeled as the user's percentile rank considering the number of citations or check-ins. To generate a query for \texttt{OAG} datasets, first, we choose 1-3 query terms from the most frequent $10,000$ author-provided keywords, and then augment each keyword with its semantically similar keywords using our \texttt{semantic keyword similarity model}. For \texttt{Gowalla} dataset, we choose a set of locations within a range of 5km as query terms.

   We vary different parameters as shown in Table \ref{table:paramList}. When one parameter is varied, other parameters are fixed at their default values.

We have uploaded the constructed attributed graphs in a public repository\footnote{https://github.com/saiful1105020/VLDB-2020-Additional-Contents}. The repository also contains a detailed description of datasets, query setting, and the \texttt{semantic similarity model} and its evaluation.

\subsection{Evaluation of semantic similarity model} \label{sec:eval_augment} 

In Section~\ref{sec:sim_model}, we presented two approaches for finding semantic similarity between two terms or keywords. Here, we empirically evaluate which approach is the most effective. 
As the ground truth, we use widely used semantic similarity measure~\cite{seco2004intrinsic} that is based on the intrinsic information content of two concepts in a given taxonomy. As the taxonomy, we use the taxonomy provided by the 2012 ACM Computing Classification System\footnote{\url{https://dl.acm.org/ccs/ccs_flat.cfm}} that contains 2,113 topics and organizes them hierarchically based on relevance (e.g., ``clustering'' is a sub-topic of ``data mining''). 
In our context, each topic in the taxonomy can be seen as a keyword or a term, and each of our proposed similarity measures can be considered as a ranker that finds the most similar topics to any topic in the taxonomy. So, we adopt Normalized Discounted Cumulative Gain ($NDCG$)~\cite{wang2013theoretical}, which is a widely accepted performance measure of ranking systems.

First, given the taxonomy $\tau$, we give the ground truth formula of the similarity between two topics $t_1$ and $t_2$ proposed in \cite{seco2004intrinsic}:
$sim_{jcn}(t_1, t_2) = 1 - [{d^{\tau}_{jcn}(t_1, t_2)}/{2}]$,
where $d^{\tau}_{jcn}(t_1, t_2)$ denotes a distance metric between $t_1$ and $t_2$ as $d^{\tau}_{jcn}(t_1, t_2)=IC^{\tau}(t_1) + IC^{\tau}(t_2) - 2 \times IC^{\tau}(lcs(t_1, t_2))$ \cite{jiang1997semantic}. Here, $lcs(t_1, t_2)$ is the “least common subsumer” in $\tau$ that subsumes $t_1$ and $t_2$. Finally, $IC^{\tau}(t)$ indicates the information content of topic $t$ in $\tau$, calculated by $\log{(\frac{|sc^{\tau}(t)|+1}{|\tau|})}/\log{(\frac{1}{|\tau|})}$ as in \cite{seco2004intrinsic}. Here, $sc^{\tau}(t)$ is the set of subsumed topics of $t$ and $|\tau|$ is the total number of topics in $\tau$. 

Second, given a topic $t$ in $\tau$,  our task is to rank the $M$-top similar topics $\{w_1, w_2, \cdots, w_M\}$ in $\tau$. The ranking goodness is evaluated by $NDCG@M$ that gives more importance on the ranking of a more relevant entity than the ranking of entities with lower relevance.

To obtain a vector representation of each topic in $\tau$, we use Google's pre-trained word2vec model\footnote{https://code.google.com/archive/p/word2vec/} that includes embedding vectors for a vocabulary of 3 million words and is trained on Google News dataset covering academic research. The length of the embedding vector is 300.

Table \ref{tab:ndcg} shows the comparison among the two similarity metrics in terms of $NDCG@50, NDCG@20, NDCG@10$. We choose \texttt{indirect cosine} that outperforms \texttt{cosine} metric.

\begin{table}[t]
    	\centering
    	\scalebox{0.9}{%
    	\begin{tabular}{ |p{0.12\textwidth}|p{0.20\textwidth}|p{0.08\textwidth}| }
    	    \hline
    		Parameter& Range & Default\\
    		\hline 
    		Dataset & \texttt{OAG}, \texttt{Gowalla} & \texttt{OAG}\\
    		\hline
    		Dataset size (vertices) & $300K$, $500K$, $700K$, $900K$, $1M$ & $500K$\\
    		\hline
    		Number of keywords & 100, 250, 500, 750, 1000 & 1000\\
    		\hline
    		$\beta$ & any real value within [0.0, 1.0] & 0.60\\
    		\hline
    		$r$ & any integer within [1, 5] & 3\\
    		\hline
    		$k_{min}$ & any integer within [2, 50] & 10\\
    		\hline
    	\end{tabular}
    	}
    	\captionof{table}{Parameters for experimental analysis}
    	\label{table:paramList}
    	\vspace{-1mm}
    \end{table}

\begin{table}[ht]
\centering
\scalebox{0.8}{%
\begin{tabular}{|c|c|c|}
\hline
\textbf{Metric} & \textbf{cosine}  & \textbf{indirect cosine}  \\ \hline
NDCG@50                                                                              & 0.528 & \textbf{0.542} \\ \hline
NDCG@20                                                                              & 0.537 & \textbf{0.607} \\ \hline
NDCG@10                                                                              & 0.573 & \textbf{0.633} \\ \hline
\end{tabular}
}
\caption{Performance results of similarity metrics}
\label{tab:ndcg}
\vspace{-4mm}
\end{table}

In the \texttt{indirect cosine} similarity measure (see Section~\ref{sec:sim_model}) there is a challenge in determining a good value for $L$. To examine this, we use two measures.
The first is \emph{word coherence, $WC^L(V^t)$}, that indicates how coherent the $L$-top words in vector $V^t$ are. The more coherent it is, the better we can represent the set of keywords. We define the word coherence as the average pairwise cosine similarity of the $L$-top words. Also, in a sense, each term is a cluster containing its relevant words. So we use \textit{Davies-Bouldin Index} \cite{davies1979cluster} as the second measure that optimizes two criteria: (1) minimizing intra-distance between words and the centroid, and (2) maximizing inter-distance between keywords. Values closer to zero indicate a better clustering.
A small value of $L$ usually contains words with high coherence and better clustering, but it is tough to find similar keywords. A large value contains words with low coherence and worse clustering, but it is easier to find similar keywords. 
The desired value of $L$ should be able to find a sufficient number of similar keywords, and provide high word coherence and low Davies-Bouldin index. 
By experimental evaluation\footnote{https://github.com/saiful1105020/Shared/blob/master/sem\_eval.pdf}, we found that $L=15$ provides a nice trade-off being able to retrieve at least $10$ similar keywords given a term in $98\%$ cases. 

\subsection{Evaluation of KICQ}
The \texttt{OAG} dataset is enriched with metadata from millions of articles, and better fits the application scenarios of our study, and thus we use it as the default dataset, unless otherwise stated, to demonstrate the performance of our proposed algorithms. First, we evaluate how different parameters (Table \ref{table:paramList}) affect the efficiency and effectiveness of the competitive algorithms. Then we compare the performance of our algorithm with the state-of-the-art influential community search algorithm~\cite{li2015influential}.

\subsubsection{Performance evaluation}
In this section, we show the scalability,  sensitivity of different parameters, memory requirement, and the cohesiveness of the retrieved communities by running a wide range of experiments.


\textbf{KICQ processing time: } In this set of experiments, we evaluate and compare the runtime of our proposed algorithms.

\begin {figure}[!htbp]
    \centering
    \begin{subfigure}[t]{0.225\textwidth}
        \centering
        \includegraphics[width=\linewidth]{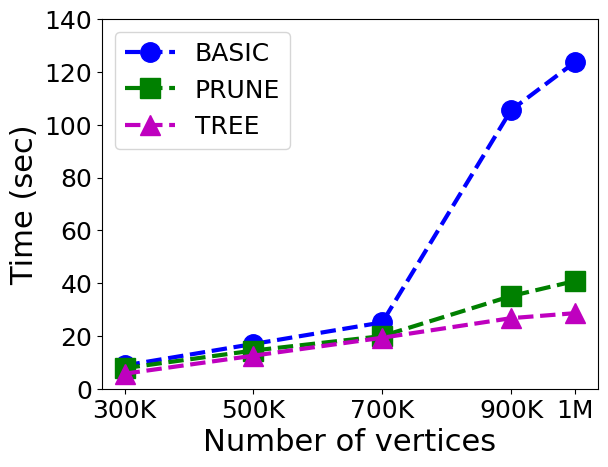}
        \caption{OR predicate}
        \label{fig:time_all_or}
    \end{subfigure}
    ~
    \begin{subfigure}[t]{0.225\textwidth}
        \centering
        \includegraphics[width=\linewidth]{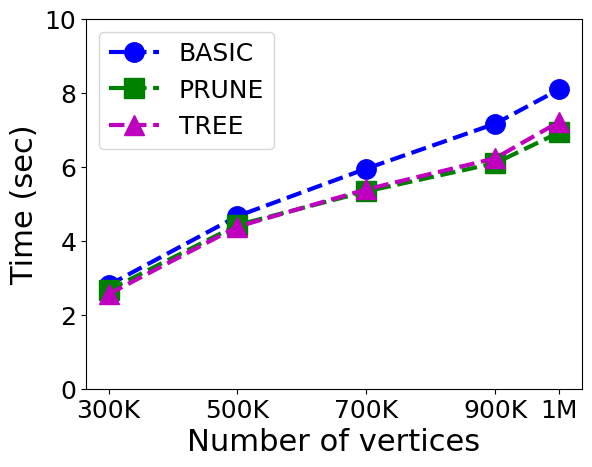}
        \caption{AND predicate}
        \label{fig:time_all_and}
    \end{subfigure}
    \caption{Query processing time for varying dataset size}
    \label{fig:runtime_scalability}
    \vspace{-4mm}
\end {figure}

\emph{Varying the dataset size.} To show the scalability, we consider different size of \texttt{OAG} dataset by varying the number of vertices and thereby edges. Figure \ref{fig:runtime_scalability} shows that with the increase in the number of vertices, runtime also increases. For AND predicate, the efficiency is nearly the same for all techniques because the query essential graph (in both \texttt{BASIC-EXPLORE} and \texttt{TREE-EXPLORE}) is small since most of the vertices are filtered out. The advantage of pruning few vertices is ruled out by the overhead of exploring the large tree in \texttt{TREE-EXPLORE}. However, for more time consuming OR predicate, \texttt{TREE-EXPLORE} is the most efficient approach, almost 1-5 times faster than the \texttt{BASIC-EXPLORE}. Also, \texttt{PRUNED-EXPLORE} and \texttt{TREE-EXPLORE} scales much better than \texttt{BASIC-EXPLORE}. 

\begin {figure}[!htb]
    \centering
    
    \begin{subfigure}{0.24\textwidth}
        \centering
        \resizebox{\textwidth}{!}{
            \includegraphics{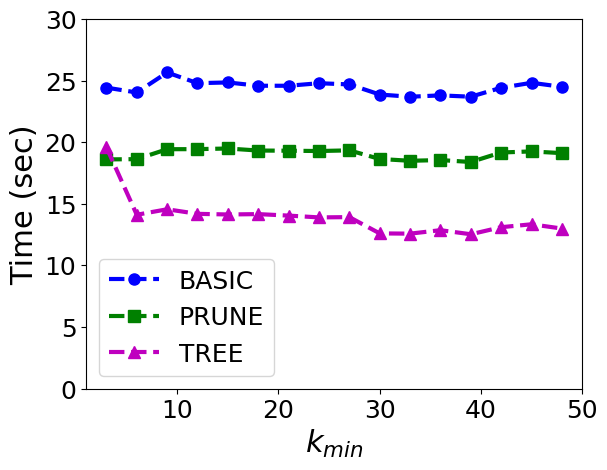}
        }
        \caption{OR predicate}
        \label{fig:kmin_time_or}
    \end{subfigure}%
    ~
    \begin{subfigure}{0.24\textwidth}
        \centering
        \resizebox{\textwidth}{!}{
            \includegraphics{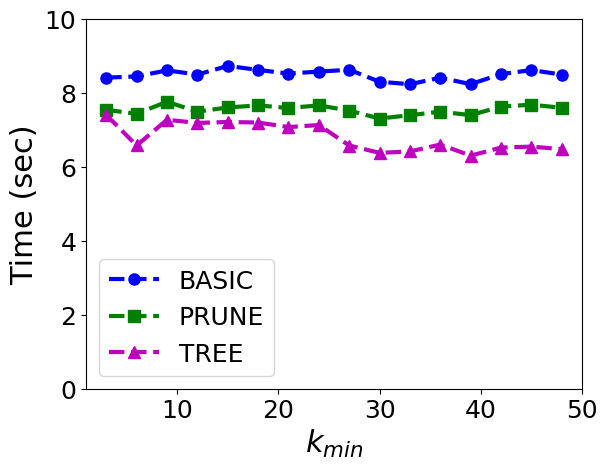}
        }
        \caption{AND predicate}
        \label{fig:kmin_time_and}
    \end{subfigure}
    \caption{Query processing time for varying $k_{min}$}
    \label{fig:runtime_kmin}
    \vspace{-2mm}
\end {figure}

\emph{Varying $k_{min}$.}\label{variableKMin} Figure \ref{fig:runtime_kmin} shows how the query processing time is affected by parameter $k_{min}$. None of the \texttt{BASIC-EXPLORE} and \texttt{PRUNED-EXPLORE} algorithms are significantly affected by the value of $k_{min}$. However, if $k_{min}$ is set to a high value, \texttt{TREE-EXPLORE} does not need to explore the tree nodes representing $k$-cores with lower $k$ values. This enables \texttt{TREE-EXPLORE} to skip a larger part of the tree since most of the vertices in the \texttt{OAG} dataset has degree less than 10 making \texttt{TREE-EXPLORE} significantly faster for higher $k_{min}$ values.

\begin {figure}[!h]
    \centering
    
    \begin{subfigure}{0.24\textwidth}
        \centering
        \resizebox{\textwidth}{!}{
            \includegraphics{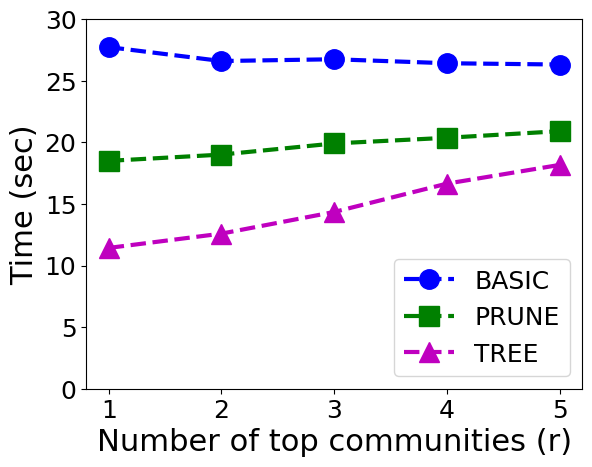}
        }
        \caption{OR predicate}
        \label{fig:r_time_or}
    \end{subfigure}%
    ~
    \begin{subfigure}{0.24\textwidth}
        \centering
        \resizebox{\textwidth}{!}{
            \includegraphics{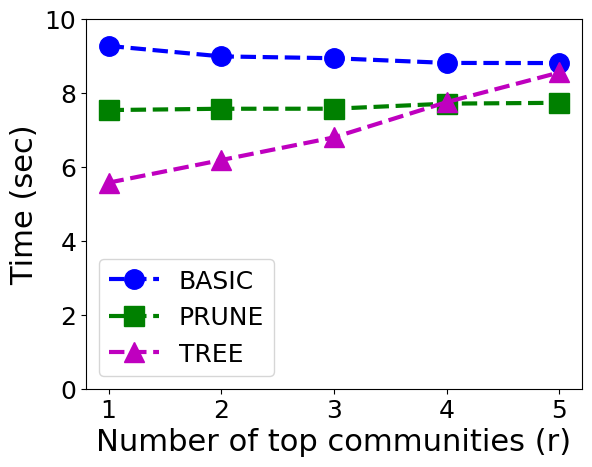}
        }
        \caption{AND predicate}
        \label{fig:r_time_and}
    \end{subfigure}
    \caption{Query processing time for varying $r$}
    \label{fig:runtime_r}
    \vspace{-2mm}
\end {figure}

\emph{Varying $r$.}\label{variableR} Figure \ref{fig:runtime_r} demonstrates how the query processing time is affected by the number of communities to be retrieved (query parameter $r$). Both \texttt{BASIC-EXPLORE} and \texttt{PRUNED-EXPLORE} compute core decomposition once on the entire query essential graph. However, \texttt{TREE-EXPLORE} performs the core decomposition on-demand basis. \texttt{BASIC-EXPLORE} is not affected by the value of $r$ since it does not use any pruning based on the retrieved communities. \texttt{PRUNED-EXPLORE} prunes some expansion based on top-$r$ score, but the query processing time is not noticeably affected. The effect is more substantial in \texttt{TREE-EXPLORE}. A significant part of the graph does not require core decomposition if $r$ is small. If $r$ is high, the algorithm can only prune a few tree nodes, but the advantage is ruled out by the overhead of decompressing the graph vertices inside a tree node. However, for small values of $r$, \texttt{TREE-EXPLORE} significantly outperforms the other algorithms, especially for OR predicate where query processing time is markedly higher than the AND predicate.

\begin {figure}[!htb]
        \centering
        
        \begin{subfigure}[t]{0.23\textwidth}
            \centering
            \includegraphics[width=\linewidth]{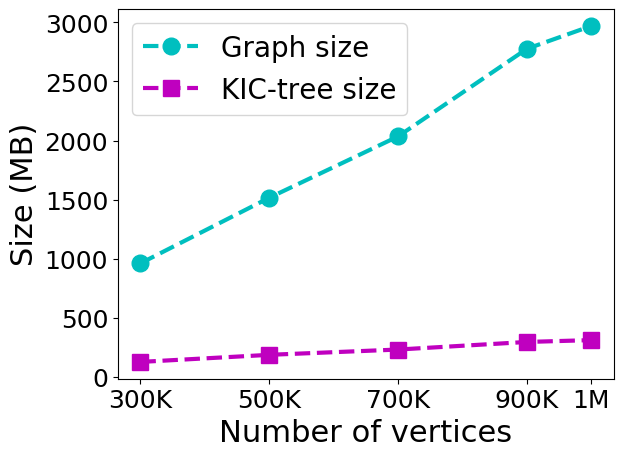}
            \label{fig:size_nodes}
            \vspace{-3mm}
        \end{subfigure}
        ~
        \begin{subfigure}[t]{0.23\textwidth}
            \centering
            \includegraphics[width=\linewidth]{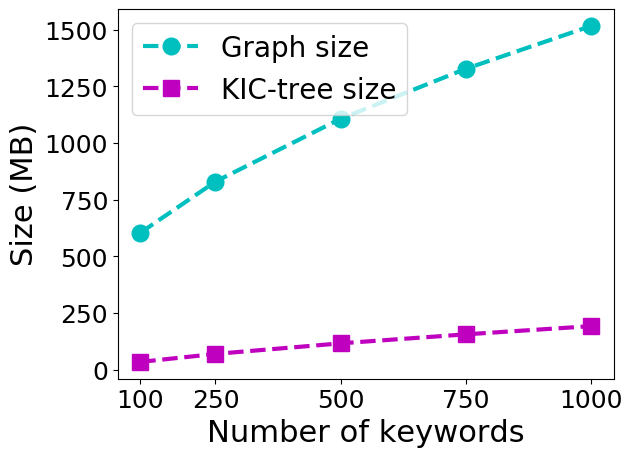}
            \label{fig:size_keywords}
            \vspace{-3mm}
        \end{subfigure}
        \caption{Index size for varying vertices and keywords}
        \label{fig:space_complexity}
        
        \vspace{-1mm}
    \end {figure}

\textbf{Index size: }
Figure \ref{fig:space_complexity} shows how the size of the graph and corresponding \texttt{KIC-tree} index increase with the increasing number of vertices and keywords. The result conforms to the complexity analysis demonstrated in Section \ref{sec:cltree_space}. Index size is linear to both the number of vertices and number of keywords if one of these remains unchanged. Also, index size is bounded by the graph size.
    
\textbf{Structural cohesiveness:} We use popular structural cohesiveness metrics diameter, density, average degree, and clustering coefficient~\cite{Fang2019} to measure the quality of the communities retrieved by our approach. These measures mostly depend on the community models (e.g., $k$-core, $k$-truss) as discussed in a survey of community search \cite{Fang2019}. They prefer the $k$-core model because of its high efficiency with minimal sacrifice in structural cohesiveness. By analyzing the cohesiveness measures in Table \ref{tab:struct_cohesiveness}, we claim that our algorithms can retrieve cohesive communities with a small diameter.

\begin{table}[!h]
\small
\scalebox{0.95}{%
\begin{tabular}{|p{0.09\textwidth}|p{0.06\textwidth}|p{0.08\textwidth}|p{0.08\textwidth}|p{0.06\textwidth}|}
\hline
Dataset       & Density & Average Degree & Clustering Coefficient & Diameter \\ \hline
\texttt{OAG}     &    $0.621$     &      $177.897$          & $0.708$ &  $2.12$  \\ \hline
\texttt{Gowalla}     &    $0.579$     &      $7.484$          & $0.881$ &  $2.70$  \\ \hline
\end{tabular}
}
\caption{Structural cohesiveness measures}
\label{tab:struct_cohesiveness}
\vspace{-3mm}
\end{table}

\begin {figure}[!htb]
    \centering
    
    \begin{subfigure}{0.23\textwidth}
        \centering
        \resizebox{\textwidth}{!}{
            \includegraphics{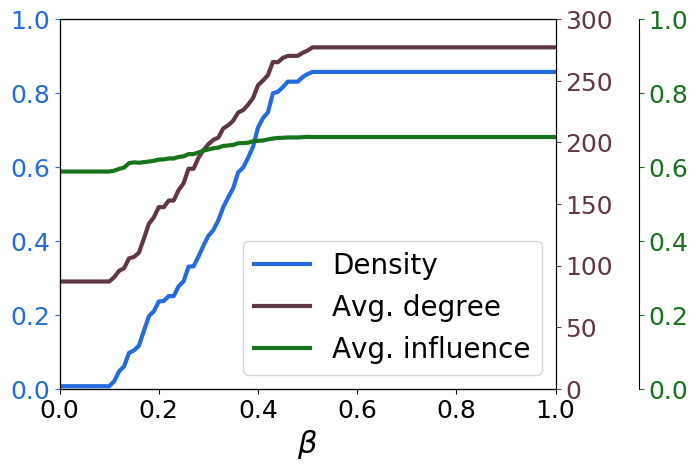}
        }
        \caption{OR predicate}
        \label{fig:beta_qual_or}
    \end{subfigure}%
    ~
    \begin{subfigure}{0.23\textwidth}
        \centering
        \resizebox{\textwidth}{!}{
            \includegraphics{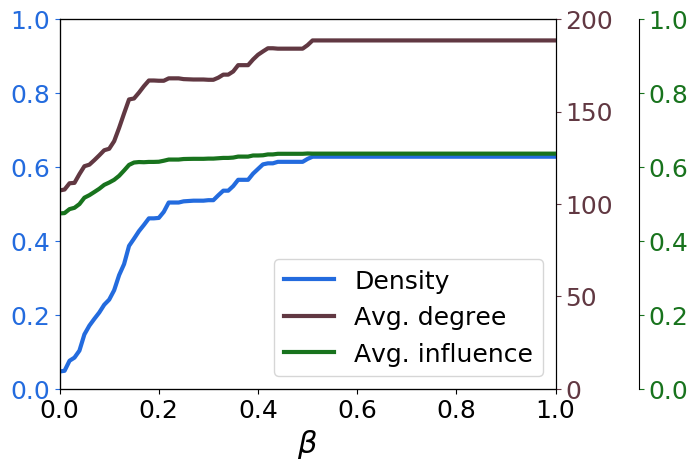}
        }
        \caption{AND predicate}
        \label{fig:beta_qual_and}
    \end{subfigure}
    \caption{Performance measures for varying $\beta$ values}
    \label{fig:beta_quality}
    \vspace{-3mm}
\end {figure}

\textbf{Setting the value of $\beta$:}\label{variableBeta}
We first run experiments to find an appropriate $\beta$, which balances the weight of connectivity and individual influence (query relevance score). For larger $\beta$, top communities are likely to show more cohesiveness and discard communities with influential individuals but less connectivity. When $\beta$ is smaller, the top communities might incline to individual influence than cohesiveness. To find a good choice of $\beta$, we consider the network structure, that is: (1)
we plot structural cohesiveness measures (i.e., density and average degree~\cite{DBLP:journals/pvldb/FangCLH16}), and (2) use the average influence score of the members for different $\beta$ values as shown in Figure \ref{fig:beta_quality}.
 
For $\texttt{OAG}$, communities with higher cohesiveness seem to contain influential individuals. This can be easily explained by the fact that an author who has co-authorship with a large number of authors is likely to have a strong influence in her field of studies. So, for $\texttt{OAG}$, we choose a high value of $\beta$ (i.e., 0.6). Note that, considering the influence of communities is still essential since it is the tie-breaker between two communities with the same cohesiveness.

\subsubsection{Comparison with state-of-the-art}

We choose \texttt{Online-All}~\cite{li2015influential} as the state-of-the-art approach as they find influential community in non-attributed graphs. 



\begin {figure}[ht]
    \centering
    
    \begin{subfigure}{0.25\textwidth}
        \centering
        \resizebox{\textwidth}{!}{
            \includegraphics{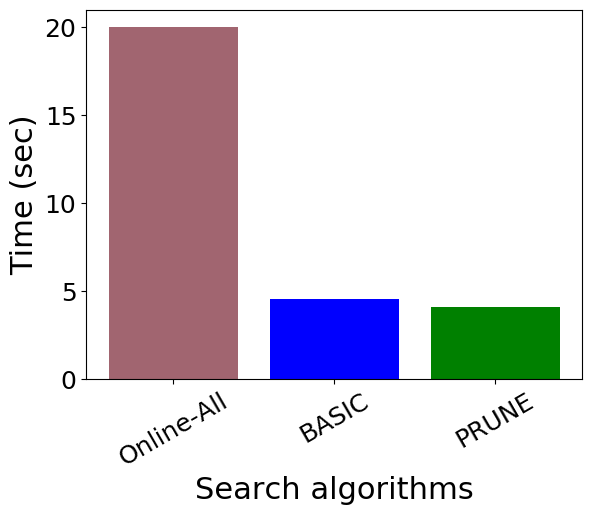}
        }
        \caption{Query processing time}
        \label{fig:runtime_all_vs_li}
    \end{subfigure}%
    ~
    \begin{subfigure}{0.25\textwidth}
        \centering
        \resizebox{\textwidth}{!}{
            \includegraphics{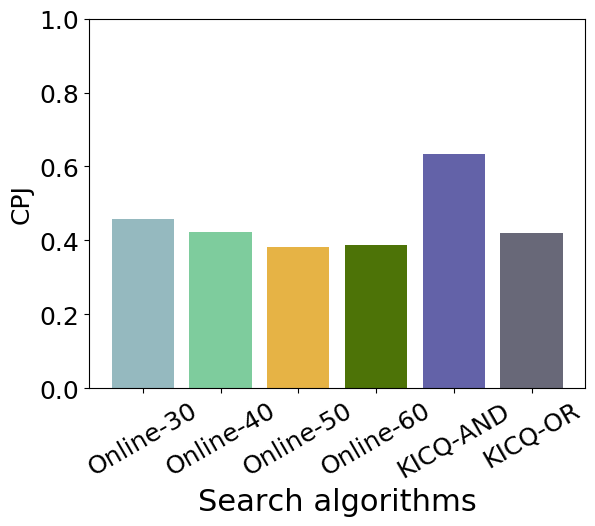}
        }
        \caption{Keyword cohesiveness}
        \label{fig:cpj_vs_li}
    \end{subfigure}
    \caption{Effectiveness and efficiency comparison}
    \label{fig:comparison}
    \vspace{-3mm}
\end {figure}

\textbf{Efficiency:}
To compare our algorithms with \texttt{Online-All}, we construct queries with a single keyword (as it does not support keywords) and compute the query essential graph, which is the input graph to \texttt{Online-All}. We also need to input the cohesiveness parameter $k$ in \texttt{Online-All}. For this, we only consider the top community ($r = 1$), and the value of $k$ in the top community returned by our approach is fed to \texttt{Online-All}. We do not consider the other algorithms in ~\cite{li2015influential} since they use pre-computation which cannot be adopted for our problem. For fair comparison, we do not consider \texttt{TREE-EXPLORE} algorithm since it uses pre-computed index. We show the query processing times of these algorithms in Figure \ref{fig:runtime_all_vs_li}. Our approaches are \emph{four times} faster compared to the \texttt{Online-All} algorithm.

\textbf{Cohesiveness:}
To compare the keyword cohesiveness, we use community pair-wise Jaccard (CPJ) metric proposed in \cite{fang2016effective}. CPJ measures the similarity of the members of top communities in terms of keywords. We form the queries as described in Section \ref{sec:setup} to evaluate our approach. \texttt{OnlineAll} cannot process queries with keywords; rather, it requires a cohesiveness parameter $k$ as input. For the simplicity of presentation, we denote our approach as KICQ-AND, KICQ-OR for AND, OR predicates, respectively. Online-$x$ denotes \texttt{OnlineAll} with parameter $k = x$. The comparison is presented in Figure \ref{fig:cpj_vs_li}. For the communities returned by our approach with AND predicate, keyword cohesiveness is $1.5$ times higher than \texttt{OnlineAll}, while for OR predicate CPJ is similar. This is expected as for \texttt{OAG}, two vertices are only connected when corresponding authors publish a paper together, and they also share common keywords. For this reason, \texttt{OnlineAll} finds communities with good CPJ value. However, this might not the case to other social networks (e.g., Gowalla, Twitter). Note that our approach and \cite{li2015influential} use the same community model (i.e., $k$-core), and the structural cohesiveness is similar. 

\subsection{Experiments with Gowalla dataset} \label{sec:exp-gowalla}
\begin {figure}[ht]
    \centering
    
    \begin{subfigure}{0.23\textwidth}
        \centering
        \resizebox{\textwidth}{!}{
            \includegraphics{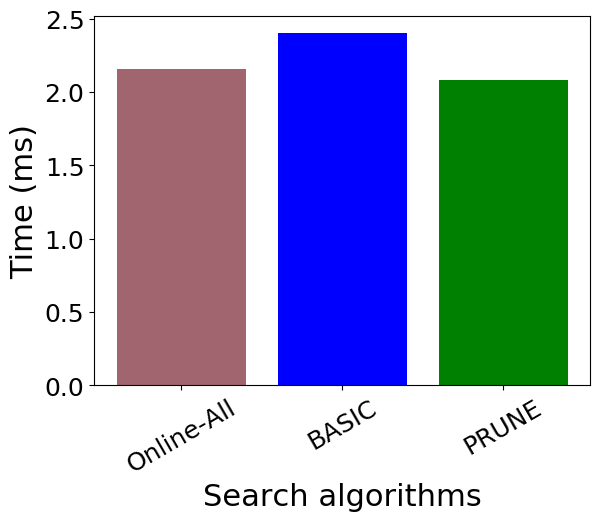}
        }
        \caption{Query processing time}
        \label{fig:gowalla_runtime_vs_li}
    \end{subfigure}%
    ~
    \begin{subfigure}{0.23\textwidth}
        \centering
        \resizebox{\textwidth}{!}{
            \includegraphics{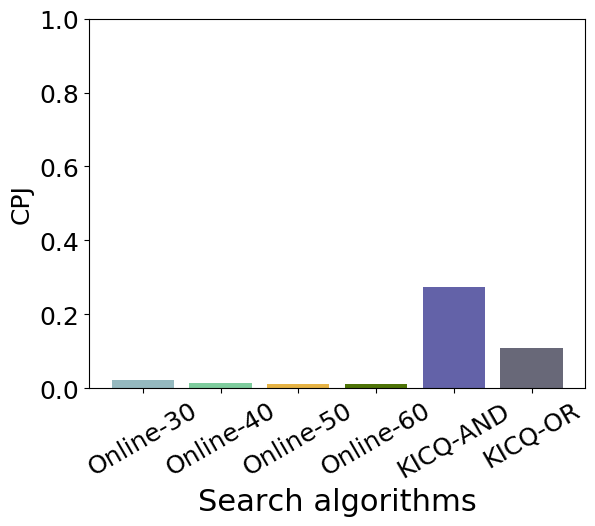}
        }
        \caption{Keyword cohesiveness}
        \label{fig:gowalla_cpj_vs_li}
    \end{subfigure}
    \caption{Effectiveness and efficiency comparison}
    \label{fig:gowalla_comparison}
    \vspace{-5mm}
\end {figure}

We conduct experiments on \texttt{Gowalla} dataset to show that our algorithms can handle large number (millions) of keywords. For these experiments, we carefully craft the queries as described in Section \ref{sec:setup}, and use $\beta = 0.5$, $r = 3$, $k_{min} = 5$ as default values. First, we report the cohesiveness measures in Table \ref{tab:struct_cohesiveness} and observe that our approach can retrieve cohesive communities with a small diameter for \texttt{Gowalla} dataset as well. We also compare the effectiveness and efficiency of the communities retrieved by our approach, and Li et al.~\cite{li2015influential}. The query setup and parameter settings are done in the same way as in the \texttt{OAG} dataset. The query processing time and keyword cohesiveness of these two approaches are presented in Figure \ref{fig:gowalla_comparison}. The number of relevant vertices is very small for queries in this dataset. The average query processing time is around 2 ms for all the algorithms. However, our key observation is that unlike \texttt{OAG}, in a dataset where connectivity is not related to keywords, the \texttt{Online-All}  fails to address the keyword cohesiveness of the communities, while our approach returns communities considering both structural and keyword cohesiveness. The results show that our approach returns communities with significantly higher (approx. 15 and 5 times for AND and OR predicate respectively) keyword cohesiveness than the \texttt{Online-All}.

\subsection{A case study}\label{sec:caseStudy}
We use a small dataset of co-author network from ArnetMiner\footnote{https://aminer.org/lab-datasets/soinf/}~\cite{tang2008arnetminer} to study the quality of retrieved communities. The dataset contains 5,411 vertices and 17,477 edges. Each vertex represents an author annotated with fields from eight different research areas:  Data Mining (DM), Web Services (WS), Bayesian Networks (BN), Web Mining (WM), Semantic Web (SW), Machine Learning (ML), Database Systems (DS), and Information Retrieval (IR), where the influence score in each field depends on the number of publications in that field. There is an edge between two authors if they publish at least two papers together.

    \begin {figure}[ht]
        \centering
        \begin{subfigure}[t]{0.23\textwidth}
            \centering
            \resizebox{\textwidth}{!}{
                \includegraphics{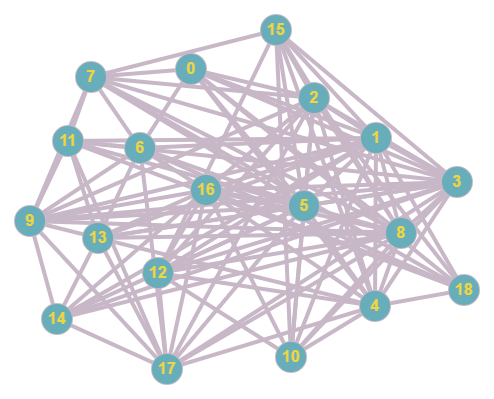}
            }
            \caption{Retrieved by our approach}
            \label{fig:top_ds}
        \end{subfigure}%
        ~    		
        \begin{subfigure}[t]{0.23\textwidth}
            \centering
            \resizebox{\textwidth}{!}{
            	\includegraphics{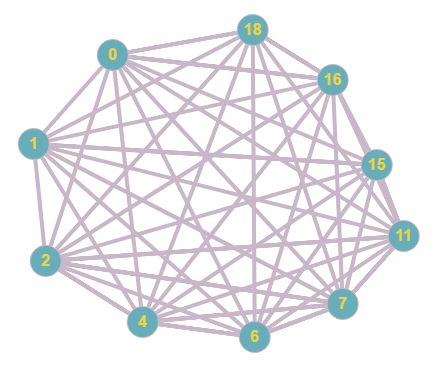}
            }
            \caption{Retrieved by \cite{li2015influential}.}
            \label{fig:top_ds_li}
        \end{subfigure}
        
        \caption{Retrieved top communities for DS.}
        \label{fig:validation_ds}
        \vspace{-3mm}
    \end {figure}

Note that \cite{li2015influential} also conducted a case study on this dataset. Figure~\ref{fig:validation_ds} shows the top community in DS retrieved by our approach (Figure~\ref{fig:top_ds}) and by \cite{li2015influential} (Figure~\ref{fig:top_ds_li}). The top community returned by our approach is 8-core and thus we compare the result of~\cite{li2015influential} for $k=8$. The details, i.e., h-index and number of citations of each author in our community are shown in Table \ref{table:top_ds}. Among them, the authors who are not included in \cite{li2015influential}'s community are shown in bold text. Due to the minimum score modelling, they missed out some of the top authors in this area including Rakesh Agrawal who was awarded the most influential scholar in the research area of Database Systems (DS) in Aminer~\footnote{https://aminer.org/mostinfluentialscholar}. Our approach keeps him in the community as we did not exclude a relatively less influential (with good connectivity) author Laura M. Haas. When Laura is not included in the community, Rakesh Agarwal is connected to less than 8 authors in the community, which turns out to be a non 8-core community. Since in the minimum weight modelling of \cite{li2015influential}, inclusion of a low influential member like Laura M. Haas significantly reduces the score of the entire community, the resultant community no longer remains the top community for $k=8$. These findings show the effectiveness of our problem formulation and score function modelling.

\begin {figure}[ht]
    \centering
    
    \begin{subfigure}{0.24\textwidth}
        \centering
        \resizebox{0.90\textwidth}{!}{
            \includegraphics{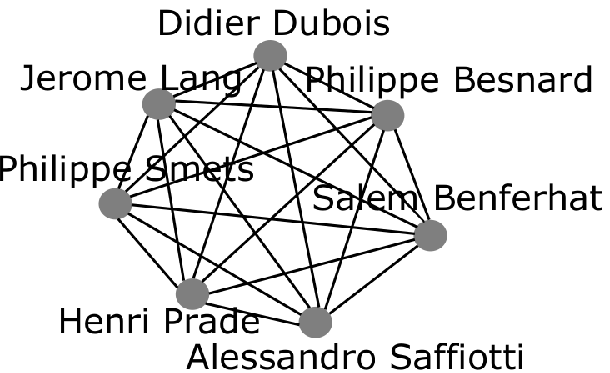}
        }
        \caption{Top most ($k=6$)}
        \label{fig:top1_bn_or_dm}
    \end{subfigure}%
    ~
    \begin{subfigure}{0.24\textwidth}
        \centering
        \resizebox{0.90\textwidth}{!}{
            \includegraphics{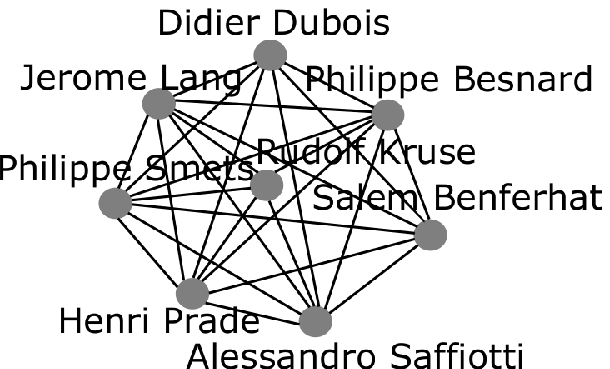}
        }
        \caption{Second top most ($k=5$)}
        \label{fig:top2_ml_and_dm}
    \end{subfigure}
    \caption{Top communities for BN OR DM.}
    \label{fig:validation_bn_or_dm}
    \vspace{-2mm}
\end {figure}

Figure \ref{fig:validation_bn_or_dm} presents two top communities for ``BN OR DM'' returned by our algorithms. The top most community is fully connected and contains highly influential authors like Didier Dubois (h-index: 125, citations: 82,295), Henri Prade (h-index: 119, citations: 78,700). The second top most community also contains all the authors from top-1 community, but the inclusion of another author Rudolf Kruse (h-index: 54, citations: 16,829) increases the contribution of individual scores, but decreases the cohesiveness of the community resulting in a lower total score than the first one. This shows the flexibility and the trade-off capability among parameters while searching for the communities. 

\begin{table}[ht]
    \centering
    \small
    \scalebox{0.90}{%
	    \begin{tabular}{ |c|c|c|c| } 
    	    \hline
    		Vertex Id & Author Name & h-index & Citations\\
    		\hline 
    		0 & Hector Garcia-Molina & 138 & 90,220\\
    		1 & David Maier & 65 & 36,687\\
    		2 & David J. DeWitt & 89 & 38,770\\
    		\textbf{3} & \textbf{Philip A. Bernstein} & \textbf{80} & \textbf{37,823}\\
    		4 & Michael Stonebraker & 72 & 26,153\\
    		\textbf{5} & \textbf{Michael J. Franklin} & \textbf{34} & \textbf{7,746}\\
    		6 & Serge Abiteboul & 80 & 35,950\\
    		7 & Jennifer Widom & 101 & 63, 641\\
    		\textbf{8} & \textbf{Joseph M. Hellerstein} & \textbf{90} & \textbf{43,207}\\
    		\textbf{9} & \textbf{Alon Y. Halevy} & \textbf{103} & \textbf{47,228}\\
    		\textbf{10} & \textbf{Jim Gray} & \textbf{81} & \textbf{46,884}\\
    		11 & Gerhard Weikum & 88 & 34,028\\
    		\textbf{12} & \textbf{Jeffrey F. Naughton} & \textbf{76} & \textbf{22,963}\\
    		\textbf{13} & \textbf{Yannis E. Ioannidis} & \textbf{59} & \textbf{15,017}\\
    		\textbf{14} & \textbf{Laura M. Haas} & \textbf{49} & \textbf{12,834}\\
    		15 & Stefano Ceri & 77 & 29,506\\
    		16 & Michael J. Carey & 59 & 16,451\\
    		\textbf{17} & \textbf{Rakesh Agrawal} & \textbf{108} & \textbf{124,595}\\
			18 & Umeshwar Dayal & 62 & 26,527\\
    				
    		\hline
    	\end{tabular}
    	}
    \captionof{table}{Top communities by our approach in DS.}
    \label{table:top_ds}
    \vspace{-5mm}
\end{table}

\section{Conclusion} \label{sec:conclusion}
In this paper, we have introduced the keyword-aware influential community query ($KICQ$) that finds top-$r$ most influential communities from an attributed graph, which has many practical applications. First, we have designed the $KICQ$ as a set of query terms conjoining with predicates (AND or OR) that enables a user to search for influential communities from an attributed graph enriched by our proposed word-embedding based keyword similarity model.  We have also proposed an influence measure for a community that considers both the cohesiveness and influence of individuals in the community. To answer the $KICQ$ efficiently, we have developed two algorithms based on the derived upper bounds and results from already explored subgraphs. Our experimental results and case study show that our approach outperforms the state-of-the-art approach in both efficiency (up to 4 times faster) and effectiveness (up to 15 times higher keyword cohesiveness). 

\balance

\newcommand{\newblock}{}
\bibliographystyle{abbrv}
\bibliography{ms}  

\end{document}